%% file: interspeech_main.tex
\documentclass[journal]{IEEEtran}

\usepackage{amsmath,graphicx}
\usepackage{cite}
\usepackage{amssymb,amsfonts}
\usepackage{algorithmic}
\usepackage{graphicx}
\usepackage{textcomp}
\usepackage{xcolor}
\usepackage{booktabs}

\begin{document}

\title{Multimodal Continuous Emotion Recognition using Deep Multi-Task Learning with Correlation Loss}
\author{Berkay K\"{o}pr\"{u}, Engin Erzin}
%The maximum number of authors in the author list is twenty. If the number of contributing authors is more than twenty, they should be listed in a footnote or in acknowledgement section, as appropriate.

\maketitle
\begin{abstract}
 In this study, we focus on continuous emotion recognition using body motion and speech signals to estimate Activation, Valence, and Dominance (AVD) attributes. Semi-End-To-End network architecture is proposed where both extracted features and raw signals are fed, and this network is trained using multi-task learning (MTL) rather than the state-of-the-art single task learning (STL). Furthermore, correlation losses, Concordance Correlation Coefficient (CCC) and Pearson Correlation Coefficient (PCC), are used as an optimization objective during the training. Experiments are conducted on CreativeIT and RECOLA database, and evaluations are performed using the CCC metric. To highlight the effect of MTL, correlation losses and multi-modality, we respectively compare the performance of MTL against STL, CCC loss against root mean square error (MSE) loss and, PCC loss, multi-modality against single modality. We observe significant performance improvements with MTL training over STL, especially for estimation of the valence. Furthermore, the CCC loss achieves more than 7\% CCC improvements on CreativeIT, and 13\% improvements on RECOLA against MSE loss.
\end{abstract}


\input{intro}

\input{method}

\input{results.tex}
\input{conclusion.tex}

\bibliography{library_o}
\bibliographystyle{IEEEtran}

\end{document}

%% file: intro.tex
\section{Introduction}

% Intro
% Emotion Recognition and importance for HCI
Human beings excite emotions in response to external and internal events of major significance to itself. The emission of emotions, also known as excitations, are represented by a 3-dimensional continuous space, and emotions like happiness and sadness correponds to regions in this space. The coordinates are Activation, Valence and Dominance (AVD) indicating activeness-passiveness, positiveness-negativeness, and dominance-submissiveness respectively\cite{avd_paper1, avd_paper2}. Emotions are episodes of coordinate changes in several components including neuro-physiologic activation, motor expression and subjective feeling but possibly also action tendencies and cognitive processes \cite{emotion_survey}.

In the context of emotion recognition, continuity of the labels separates into Continuous Emotion Recognition (CER) and Emotion Classification (EC). CER research focuses on prediction of AVD values \cite{narjis_multimodal,mse_concordance_loss_end_to_end_multi_modal_schuller}, while EC focuses on classification of discrete emotions \cite{feature_discrete, feature_cnn_discrete_er, discrete_er2, discrete_er3}. Another significant difference between CER and EC studies is the loss function that is used during training. While CER studies are mandated by mean squared error (MSE) \cite{narjis_multimodal}, EC is mandated by cross-entropy \cite{discrete_er3}.

% Speech, Body M and multi-modal
%
CER research initially concentrated on the speech, and produced uni-modal regressors as speech not only highly reflects the neuro-physiological changes in the research but also its maturation in terms of analysis \cite{speech1_cer_1999,speech2_2003, correlation_loss_schuller_2019}. However, studies were recently directed into multi-modality where descriptors from body, face, speech and text are combined \cite{narjis_multimodal, mtl_mse_multi_modal_2017}. Especially, combining speech with body motion and facial expressions has become a common practice, and multi-modal approaches have quickly proven their superiority against the uni-modal approaches \cite{narjis_multimodal, mtl_mse_multi_modal_2017}.

% Deep learning vs traditional learning
Traditionally, CER set-up consists of a feature extraction, feature summarization and regression blocks  \cite{narjis_multimodal, feature_ann1, feature_discrete}. With the rise of Deep Neural Networks (DNN) and their ability to model non-linear behaviors, initially the Feed-Forward Networks occupied the regressor blocks \cite{feature_ann1}. Later, Recurrent Neural Networks (RNN) and Convolutional Neural Networks (CNN) were introduced into CER. RNNs replaced the Feed-Forward Networks to capture long term relationships \cite{feature_based_cer_rnn}, while CNNs were adapted successfully as a meta-feature extractor by \cite{receptive_field_cnn, feature_cnn_discrete_er, feature_cnn_cer}. 
%End-to-end, semi-end-to end
As extracting low level descriptors such as acoustic features requires domain expertise and suffers from task independence, end-to-end learning, where raw signals are fed into deep architectures, were proposed as the replacement \cite{end_end_first_paper}. Such end-to-end systems then successfully adapted to the emotion recognition problem by \cite{mse_concordance_loss_end_to_end_schuller, mse_concordance_loss_end_to_end_multi_modal_schuller, main_paper_schuller2019}.

% Multi-task learning,
The state-of-art approaches attack the CER problem by training a predictor for each output. This approach not only increases the complexity of the general task but also suffers from over-fitting due to limited amount of training data. By training the predictor with multi-task learning (MTL), one can force it to learn shared representations and increase the generalization ability \cite{theory_mtl}. Although MTL is exploited by \cite{main_paper_schuller2019, mtl_mse_speech_only_2017, mtl_mse_multi_modal_2017} in the context of CER, unlike our approach these studies trained the proposed architectures with the MSE loss.

% Correlation Loss % ADD general losses used in literature
Studies on CER evaluate themselves regarding correlation metrics Concordance Correlation Coefficient (CCC) \cite{correlation_loss_schuller_2019}, or Pearson Correlation Coefficient (PCC) \cite{narjis_multimodal} while training the predictors using different type of error metrics such as MSE or mean absolute error (MAE) loss . This mismatch causes a degradation in the performance of proposed architectures as also discussed in \cite{mse_concordance_loss_end_to_end_schuller, mse_concordance_loss_end_to_end_multi_modal_schuller}. 

% contributions
In response to these issues in the research, our study tries to address these weaknesses. The main contributions of this study are below:
\begin{itemize}
	\item A multi-modal deep neural network architecture based on speech and body motion is proposed.
	\item An  MTL based architecture surpassing single task learning (STL) state-of-art
	\item A correlation loss metric that is used as objective function during the training
\end{itemize}

To the best of our knowledge, this is the first study investigating semi end-to-end framework exploiting MTL and correlation loss in the context of multi-modal CER.

%% file: method.tex
\section{Methodology} \label{sec:method}
In this section, we first define the feature representations of the speech and body motion signals. Then, the dataset that the experiments are conducted on is introduced, and finally we provide the details of the proposed framework.

\subsection{Feature Representations}
Mel-frequency cepstral coefficients (MFCC) are extracted from the speech signals, and utilized to generate the acoustic features. Each speech frame is summarized into the 39 dimensional acoustic feature vector which includes the energy, the first 12 MFCCs and their first and second derivatives. The MFCC feature vector is represented as $f^{s}_{l} \in  \mathbb{R}^{39\times1}$ for the $l$th time frame.

To represent the body motion, at each instance raw Euler rotation angles in (x,y,z) dimensions along with their deltas are utilized. The resulting body motion feature vector for the $l$th frame is defined as $f^{b}_{l} \in \mathbb{R}^{24\times 1}$.

\subsection{Dataset}
In this study, USC CreativeIT \cite{database_creative_it_busso_2016} and RECOLA databases are used to train the proposed architecture. The CreativeIT database consist of pairs of 3.5~minutes length interplays performed by 16 actors to mimic dyadic interactions. After removal of recordings that have missing labels for any of AVD, the training dataset is generated from 40 recordings. Then the recordings are divided into 5 sessions regarding mutual exclusiveness of the speakers. Dividing the dataset into this mutually exclusive sessions enables speaker independent CER. 

\begin{figure}[htbp]
\centerline{\includegraphics[width=\columnwidth]{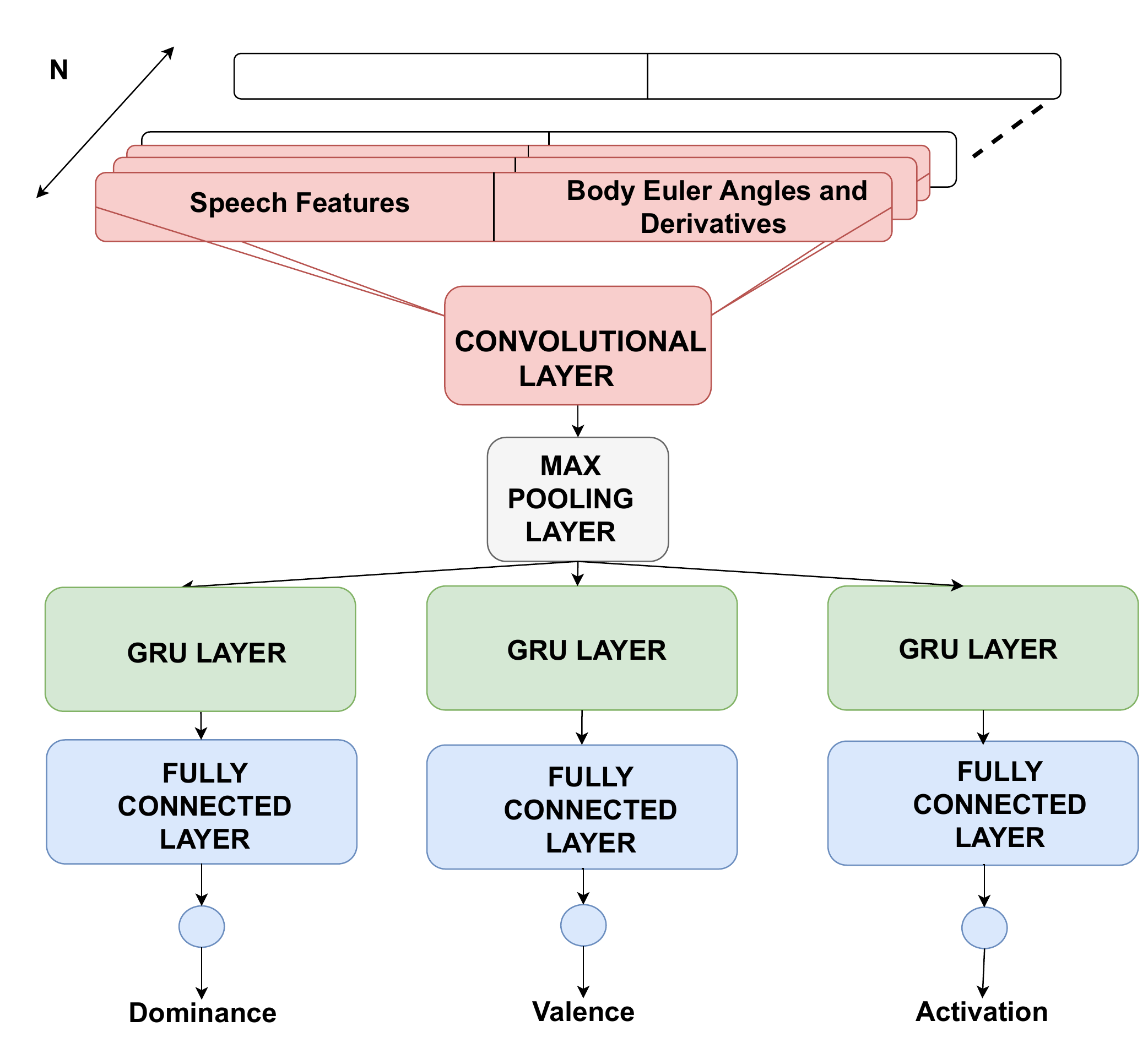}}
\caption{Semi-End-to-End framework exploiting MTL for CER}
\label{fig:architecture}
\end{figure}

The RECOLA database is a popular database for audiovisual emotion recognition, as it is used for the AVEC challenges \cite{avec2015}, \cite{avec2016} and \cite{avec2018}. The database contains dyadic interactions of 27 French-speaking subjects. The recordings are separated into three parts 9 train subjects, 9 development and 9 test subjects, as the AVEC challenges urge. By following the procedure in \cite{recolaValOnlyWu2019}, only train set is used for training and results are provided from development set. The database is annotated with frame rate of 40 ms, by six annotators. In this study, the mean of these 6 annotations are used as the reference. In addition, while experimenting on RECOLA, body Euler angles in Figure~\ref{fig:architecture}, are replaced with the facial activation units, yaw, pitch, mean and standard deviation of optical flow.

\subsection{Framing} 
In order to capture the temporal continuity and variations in the features and to consider the slow varying nature of emotions, we choose a sequence of feature vectors to form a temporal feature-image representation. Since the frame rate of the speech and body features is 60~fps, we choose to form the sequence of feature vectors with a stride of $t=10$ frames as
\begin{equation}
    F_{l} = [f_{l+1-Nt/2},\dots,f_{l-t}, f_l, f_{l+t},f_{l+2t},\dots,f_{l+Nt/2}] 
\end{equation}
where $F_l \in \mathbb{R}^{P\times N}$ is the feature image at frame $l$, $P$ is the feature vector dimension and $N$ is the number of frames defining the temporal extent.

\subsection{Architecture}
The proposed semi-end-to-end architecture is depicted in the Figure~\ref{fig:architecture}. The proposed architecture mimics a single-task end-to-end framework which is composed of a multiple CNN layers, Max-Pooling Layers, RNN layers and a final fully connected layer \cite{mse_concordance_loss_end_to_end_schuller}. In this architecture, CNN acts as a meta-feature extractor having the ability to extract task-dependent features or meta-features either from the raw signal or input features.

RNN's are famous for their ability to model sequential information. Combination of CNN and RNN layers are effectively used to find the long-term relationships for the image captioning task \cite{dnn_rnn_image_caption1, dnn_rnn_image_caption2}. 
\begin{figure*}[!h]
\centerline{\includegraphics[width=\paperwidth]{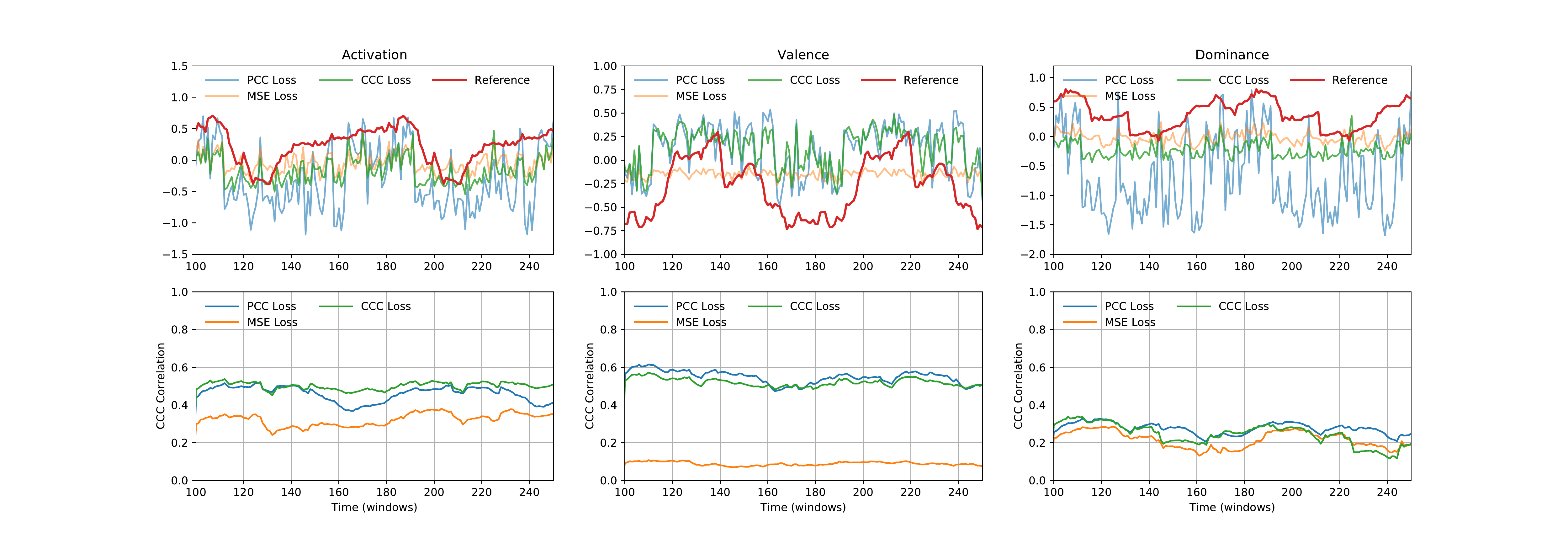}}
\caption{Prediction of AVD modalities over time for correlation loss and MSE loss}
\label{fig:corr_rmse_time_plot}
\end{figure*}
On top of the single end-to-end framework (CER-STL), the proposed architecture involves MTL and these tasks share the same meta-feature extractor. We will refer to the proposed architecture as CER-MTL.

\subsubsection{Modality Fusion}
In our framework, we apply an early fusion and concatenate the modalities at the input. Hence, the input of the multimodal architecture is $F^{b,s}_{l} \in \mathbb{R}^{63\times N}$. Concatenating the modalities at the input is both beneficial in terms of complexity of the total architecture, and also from the learning point of view. The most complexity hungry part of the proposed architecture is the CNN layer due to the large input dimensionality. As a result, using the same layer for each task benefits from complexity. In addition, early fusion enables the learning of cross modal features. 

\subsubsection{Loss}
In this study we utilize the PCC and CCC as loss functions, in mutually exclusive setups, between the batch of ground truth and predictions.
%\begin{gather}
%   L_{PCC}(\mathbf{y},\mathbf{\hat{y}}) =  \frac{\sum_{i=1}^{K}{(y_{i}-\mu_{y})(\hat{y}_{i}-\mu_{\hat{y}})}}{\sqrt{\sum_{i=1}^{K}{(y_{i}-\mu_{y})^2 (\hat{y}_{i}-\mu_{\hat{y}})^2}}} \\
%    L_{CCC}(\mathbf{y},\mathbf{\hat{y}}) =  \frac{2\sum_{i=1}^{K}{(y_{i}-\mu_{y})(\hat{y}_{i}-\mu_{\hat{y}})}}{\sum_{i=1}^{K}{(y_{i}-\mu_{y})^2+\sum_{i=1}^{K}{(\hat{y}_{i}-\mu_{\hat{y}})^2+(\mu_{y}-\mu_{\hat{y}})^2}}}
%\end{gather}

\subsubsection{Multiple Task Learning}
The objective function of the MTL could be written as the weighted sum of objective functions of individual task which are predicting activation, valence or dominance. Let $J(\theta)$ be the objective function of MTL given as
\begin{equation}
    J(\theta) = \sum_{m=1}^{M} {\alpha_{m}L_{m}},
\end{equation}
where $\alpha_m$ is the weight of the loss of the $m^{\text{th}}$ task $L_{m}$ where $m\in\{\text{activation},\text{valence},\text{dominance}\}$.

%% file: results.tex
\section{Experiments and Results}
The proposed architecture is implemented in Keras framework with TensorfFlow backend. All the conducted experiments are held on NVIDIA GeForce GTx 1080 Ti. During the training Adam optimizer with a learning rate of 0.00005 is used. The training is held with a batch size of 256 for 100 epochs with an early stopping option of 20 epoch. All these hyper-parameters are optimized during the experiments. 

The CNN layer of the CER networks has 25 filters with kernel size of 3, and the max-pooling layer has a kernel size of 2. The output features are then fed into 5 GRU nodes per task. To add further non-linearity the GRU's outputs are fed into a fully connected layer having 2 node, and finally a linear node is added to regress the AVD values. The weights of the MTL are selected uniform as $\alpha_{m} = 1/3$ for $m \in \{1,2,3\}$.

To depict the effectiveness of the proposed architecture we devised experiments comparing MTL with STL, correlation losses (PCC and CCC) with MSE loss, multi-modality with single modality and different window sizes. For all the experiments held, CCC is used as the evaluation metric.

\begin{table*}[htb]
\begin{center}
\caption{Mean CCC comparison between CCC loss, PCC loss and MSE loss on CreativeIT database for unimodal and multimodal settings ($N=20$, temporal extent is $Nt=3.3$~sec)}
    \label{tab:corr_mse}

\begin{tabular}{l|c c c |c c c |c c c}
\toprule[1pt]\midrule[0.3pt]
%\cline{2-7}
\textbf{Model} &\multicolumn{3}{c|}{\textbf{CCC Loss}} & \multicolumn{3}{c|}{\textbf{PCC Loss}}    & \multicolumn{3}{c}{\textbf{MSE Loss}}    \\ 
       & Act & Val & Dom & Act & Val & Dom & Act & Val & Dom \\ \midrule \hline
\textbf{Multimodal} & 0.38 & 0.16 & 0.18 & 0.33       & 0.16    & 0.14      & 0.26       & 0.07    & 0.11      \\ \hline
\textbf{Motion}      & 0.15 & 0.06 &0.10 & 0.14       & 0.02    & 0.09      & 0.12       & 0       & 0.04      \\ \hline
\textbf{Speech}      & 0.37 & 0.13 &0.18 & 0.32       & 0.13    & 0.18      & 0.23       & 0.08    & 0.06      \\ \hline
\end{tabular}
\end{center}
\end{table*}

\begin{table*}[htb]
\begin{center}
\caption{Mean CCC comparison between CCC loss, PCC loss and MSE loss on RECOLA database for unimodal and multimodal settings ($N=20$, temporal extent is $Nt=3.3$~sec)}
\label{tab:corr_mse2}

\begin{tabular}{l|c c |c c |c c}
\toprule[1pt]\midrule[0.3pt]
%\cline{2-7}
\textbf{Model} &\multicolumn{2}{c|}{\textbf{CCC Loss}} & \multicolumn{2}{c|}{\textbf{PCC Loss}}    & \multicolumn{2}{c}{\textbf{MSE Loss}}    \\ 
       & Act & Val & Act & Val & Act & Val \\ \midrule \hline
\textbf{Multimodal} & 0.66 & 0.34   & 0.49 & 0.24       & 0.50 & 0.21        \\ \hline
\textbf{Facial Attributes}      & 0.15 & 0.06   & 0.09 & 0.18       & 0.22 & 0.19         \\ \hline
\textbf{Speech}      & 0.61 & 0.32   & 0.34 & 0.13       & 0.49 & 0.19        \\ \hline
\end{tabular}
\end{center}
\end{table*}

Results in Table~\ref{tab:corr_mse} demonstrate the effect of loss function and using different modalities for CER. For the multimodal case, CER-MTL with CCC loss outperforms the same architecture with PCC and MSE losses, while PCC loss outperforms MSE loss for all the of prediction modalities. For instance the proposed framework achieves 0.38 CCC for activation when it is trained with CCC loss, the same architecture achieves only 0.33 and 0.26 CCC's when it is trained with respectively PCC and MSE losses. These results state that using correlation based metrics are more suitable as the loss function when the evaluation metrics is also correlation based. In fact, with the results presented in Table~\ref{tab:corr_mse2} where CCC loss achieves at least 10\% improvements, PCC loss fails to outperform MSE loss on prediction of activation. With regard to this observation, our argument on correlation metrics and losses can be strengthen as, using evaluation metric as the loss function of the DNN architecture provides a robust performance enhancement. However, the downfall of the correlation based loss functions are they require longer batch sizes than regular error metrics like MSE. As the longer batch sizes prone to over-fitting the learning rates should be chosen carefully. In our study, we used relatively small learning rates like between 0.0001 and 0.00005. However, our findings regarding over-fitting and learning rate are aligned with \cite{main_paper_schuller2019}.

The Figure~\ref{fig:corr_rmse_time_plot} depicts the predicted AVD values and the CCC correlation of the predictions from CER-MTL with CCC loss, PCC loss and MSE loss over 256 frames, where the CCC values at time index $l$ is calculated by considering the previous 100 predictions before time $l$. The proposed CCC training outperforms the MSE training by at least $10\%$ CCC difference. In addition, MSE training is unresponsive to the significant changes in the reference, while CCC and PCC loss is highly responsive. 

The performance comparison of unimodal and multimodal architectures is presented in Table~\ref{tab:corr_mse}. Our experiments show that speech carry more information about affective states as it outperforms the motion based architecture with at least 10\% CCC difference on creative IT and 2\% CCC on RECOLA. Especially, we found that body motion is almost uncorrelated with valence attribute. Another observation in regardless of the loss function activation and valence attributes are mandated by the speech modality.

\begin{table}[htb]
\begin{center}
\caption{Mean CCC comparison between CER-MTL and CER-STL multimodal architecture CCC loss with $N=20$ (temporal extent is $Nt=3.3$~sec)}
\label{tab:mtl_stl}
    \begin{tabular}{l|c c c |c c}
    \toprule[1pt]\midrule[0.3pt]
    \textbf{Model} &\multicolumn{3}{c|}{\textbf{CreativeIT}} & \multicolumn{2}{c}{\textbf{RECOLA}}\\
 & Act & Val & Dom & Act & Val \\ \midrule \hline
{MTL}   & {0.38}       & {0.16}    & {0.18}  & {0.66} & {0.34}\\ \hline
{STL}   & {0.31}       & {0.10}    & {0.16} & {0.62} & {0.27}\\ \hline
\end{tabular}
\end{center}
\end{table}
The CCC results of the proposed MTL system and the baseline STL system are given in Table~\ref{tab:mtl_stl}. CER-MTL outperforms CER-STL on predicting activation and valence attributes. Especially, MTL brings 6\% CCC improvement for valence. From Table~\ref{tab:mtl_stl} one can deduce that learning features that are significant for both activation and dominance helps predicting the valence. MTL improves the results not only by creating a regularization effect for the other tasks, but also for the special case of CER, it exploits the correlation between the AVD values. Another benefit of the MTL arises in terms of complexity and resource efficiency. Baseline results are generated by training CER-STL for each task. Training time and the total complexity of the networks for the baseline is almost multiple of the total number of task. 

Table~\ref{tab:ws_mtl_corr} demonstrates the effect of using different window sizes for CER. Note that increasing the window size significantly effect the prediction of activation while there is no such significant trend for valence and dominance. Intuitively increasing receptive field size would add more information to be learnt, the slow varying nature of affective states might add repetition rather than information. Hence, the increasing the input size directly effects the complexity/number of parameters of the CER-MTL which makes it prone to over-fitting.  
\begin{table}[!h]
\begin{center}
\caption{Effect of temporal extent ($N$) on CER for the MTL model with the CCC loss}
\label{tab:ws_mtl_corr}
\begin{tabular}{c|c c c |c c}
\toprule[1pt]\midrule[0.3pt]
\textbf{N} &\multicolumn{3}{c|}{\textbf{CreativeIT}} & \multicolumn{2}{c}{\textbf{RECOLA}}\\
 & Act & Val & Dom & Act & Val \\ \midrule \hline
{20} & {0.38}       & {0.16}    & {0.18}  &{0.66}   &{0.34} \\ \hline
{40} & {0.42}       & {0.16}    & {0.18}  & {0.74}  &{0.35} \\ \hline
{60} & {0.45}       & {0.16}    & {0.18}    &{0.79} &{0.37}\\ \hline
\end{tabular}
\end{center}
\end{table}

In \cite{correlation_loss_schuller_2019} the efficiency of the RNN's are discussed, and as a conclusion a complex enough CNN is found enough to model emotions via speech. We have conducted the similar experiments in this study and results are depicted in Table~\ref{tab:rnn_effect}. We compared our architecture (CER-MTL) with the pure CNN architecture from \cite{correlation_loss_schuller_2019} with a modification at the filter lengths due to the size of data. The original CER-MTL surpasses the one proposed in \cite{correlation_loss_schuller_2019} by at least 11\% CCC. Although originally Pure CNN \cite{correlation_loss_schuller_2019}, trained on SEWA corpus\cite{sewa2019}, the similar performance results are expected on the CreativeIT database as well since it includes 120~minutes of training set. This duration is comparable to size which is 130~minutes of training and development duration in \cite{correlation_loss_schuller_2019}. The performance degradation of Pure CNN \cite{correlation_loss_schuller_2019}, could be due to its high complexity, which cause overfitting.

Table~\ref{tab:recola_best_performance} presents the performance comparison of the proposed CER-MTL architecture with the state-of-the-art solutions on the RECOLA dataset. CER-MTL ($N=60$) outperforms the state-of-the-art solutions by at least 1\%~CCC on activation prediction. This comparison represents the power of the proposed architecture. 
\begin{table}[!h]
\begin{center}
\caption{Comparison of CCC performances and the number of parameters of the architectures on the CreativeIT database}
\label{tab:rnn_effect}
\begin{tabular}{l c c c c}
\toprule[1pt]\midrule[0.3pt]
\textbf{Model}  & \textbf{Act} & \textbf{Val} & \textbf{Dom} & \textbf{Param} \\ \midrule \hline
{Pure CNN \cite{correlation_loss_schuller_2019}}    & {0.15}       & {0.05}    & {0.05}      &   240,643   \\ \hline
{CER-MTL}     & {0.38}       & {0.16}    & {0.18}      &   5,615\\ \hline
\end{tabular}
\end{center}
\end{table}

\begin{table}[!h]
\begin{center}
\caption{Comparison of CCC performances on the development set of RECOLA database}
\label{tab:recola_best_performance}
\begin{tabular}{l c c}
\toprule[1pt]\midrule[0.3pt]
\textbf{Model}  & \textbf{Act} & \textbf{Val} \\ \midrule \hline
{Pure CNN \cite{correlation_loss_schuller_2019}}    & {0.51}       & {0.26}   \\ \hline
{End-to-end \cite{trigeorgisSchuller2016adieuEnd2EndRecola}} & {0.74} & {0.33}\\ \hline
{DDAT \cite{zhangSchuller2018dynamicRecola}} & {0.78} & {0.50}\\ \hline
{End-to-end Multimodal \cite{tzirakisSchuller2017end2endRecola}} &{0.75} &{0.41}\\ \hline
{FER-P\&G-Net \cite{recolaValOnlyWu2019}} &{0.60} &{0.69} \\ \hline
\textbf{CER-MTL ($N=60$)}     & \textbf{0.79}       & {0.37}\\ \hline
\end{tabular}
\end{center}
\end{table}

%\begin{figure}[!h]
%\centerline{\includegraphics[width=7cm]{cnn_only_architectures.pdf}}
%\caption{CNN based architectures left (CER-MTL CNN) and right (Pure-CNN %\cite{correlation_loss_schuller_2019}) compared with CER-MTL}
%\label{fig:cnn_mtl_architecture}
%\end{figure}

%% file: conclusion.tex
\section{Conclusion}

The proposed CER-MTL framework combines the early fusion of speech and body motion modalities, MTL and correlation loss. MTL both increases the robustness of the training and performance by forcing to learn shared representations. Moreover, early fusion and MTL bring significant saving in the complexity of the architecture that prevents the overfitting with the limited affective labeled data.

We conducted experiments on CreativeIT and RECOLA databases to capture effect of window size on the CER, MTL and CCC loss. We found that increasing window size brings a significant improvement on the prediction of activation. The results depict that CER-MTL outperforms CER-STL, especially brings a 6\% CCC improvement on prediction of valence. In addition, experiments showed that CER-MTL achieves at least 7\% and 13\% higher CCC values on respectively CreativeIT and RECOLA datasets when the models are trained with the CCC loss rather than the MSE loss. These performance improvements suggest the effectiveness of the proposed system.